\documentclass{article}

    \PassOptionsToPackage{numbers, compress}{natbib}

    \usepackage[preprint]{neurips_2024}

\usepackage[utf8]{inputenc} 
\usepackage[T1]{fontenc}    
\usepackage{hyperref}       
\usepackage{url}            
\usepackage{booktabs}       
\usepackage{amsfonts}       
\usepackage{nicefrac}       
\usepackage{microtype}      
\usepackage{xcolor}         

\usepackage{graphicx}
\usepackage{amsmath}
\usepackage{xspace}
\usepackage{caption}
\usepackage{subcaption}
\usepackage{wrapfig}

\def\enface{\emph{en~face}}
\def\Enface{\emph{En~face}}
\def\eg{e.g., } 
\def\ie{i.e., } 

\title{Conditioning 3D Diffusion Models with 2D Images: Towards Standardized OCT Volumes through \emph{En~Face}-Informed Super-Resolution}

\author{%
Coen de Vente$^{1,2,3}$ \quad Mohammad Mohaiminul Islam$^{1,2}$ \quad \textbf{Philippe Valmaggia}$^{4,5}$ \\ \textbf{Carel Hoyng}$^{6}$ \quad \textbf{Adnan Tufail}$^{7}$
\quad \textbf{Clara I. Sánchez}$^{1,2}$ \\ on behalf of the MACUSTAR consortium\thanks{The list of MACUSTAR consortium members is in Appendix Section \ref{sec:diffusion:appendix:macustar_consortium}.} \\
$^1$qurAI Group, Informatics Institute, University of Amsterdam, The Netherlands \\ $^2$Amsterdam UMC location University of Amsterdam, Biomedical Engineering and Physics,\\ The Netherlands \\
$^3$DIAG, Department of Radiology and Nuclear Medicine, Radboudumc, The Netherlands \\
$^4$Department of Biomedical Engineering, Universität Basel, Switzerland \\
$^5$Department of Ophthalmology, University Hospital Basel, Switzerland \\
$^6$Department of Ophthalmology, Radboudumc, Nijmegen, The Netherlands \\
$^7$Moorfields Eye Hospital NHS Foundation Trust, London, United Kingdom \\ 
\texttt{\{c.w.devente,m.islam,c.i.sanchezgutierrez\}@uva.nl}\\
\texttt{philippe.valmaggia@unibas.ch} \quad
\texttt{carel.hoyng@radboudumc.nl} \\
\texttt{adnan.tufail@nhs.net}
}

\begin{document}

\maketitle

\begin{abstract}
High anisotropy in volumetric medical images can lead to the inconsistent quantification of anatomical and pathological structures. Particularly in optical coherence tomography (OCT), slice spacing can substantially vary across and within datasets, studies, and clinical practices. We propose to standardize OCT volumes to less anisotropic volumes by conditioning 3D diffusion models with \enface{} scanning laser ophthalmoscopy (SLO) imaging data, a 2D modality already commonly available in clinical practice. 
We trained and evaluated on data from the multicenter and multimodal MACUSTAR study.
While upsampling the number of slices by a factor of 8, our method outperforms tricubic interpolation and diffusion models without \enface{} conditioning in terms of perceptual similarity metrics.
Qualitative results demonstrate improved coherence and structural similarity. Our approach allows for better informed generative decisions, potentially reducing hallucinations.
We hope this work will provide the next step towards standardized high-quality volumetric imaging, enabling more consistent quantifications.
\end{abstract}


\section{Introduction}

Volumetric medical images can be highly anisotropic, \ie having high-resolution slices in one anatomical plane but poor through-plane resolution. 
This has been shown to lead to imprecise volume and shape measurements of structures of interest \cite{Muld19}, potentially resulting in wrong diagnoses and severe negative clinical implications.

\begin{figure*}[!t]
    \centering
    \includegraphics[width=\linewidth]{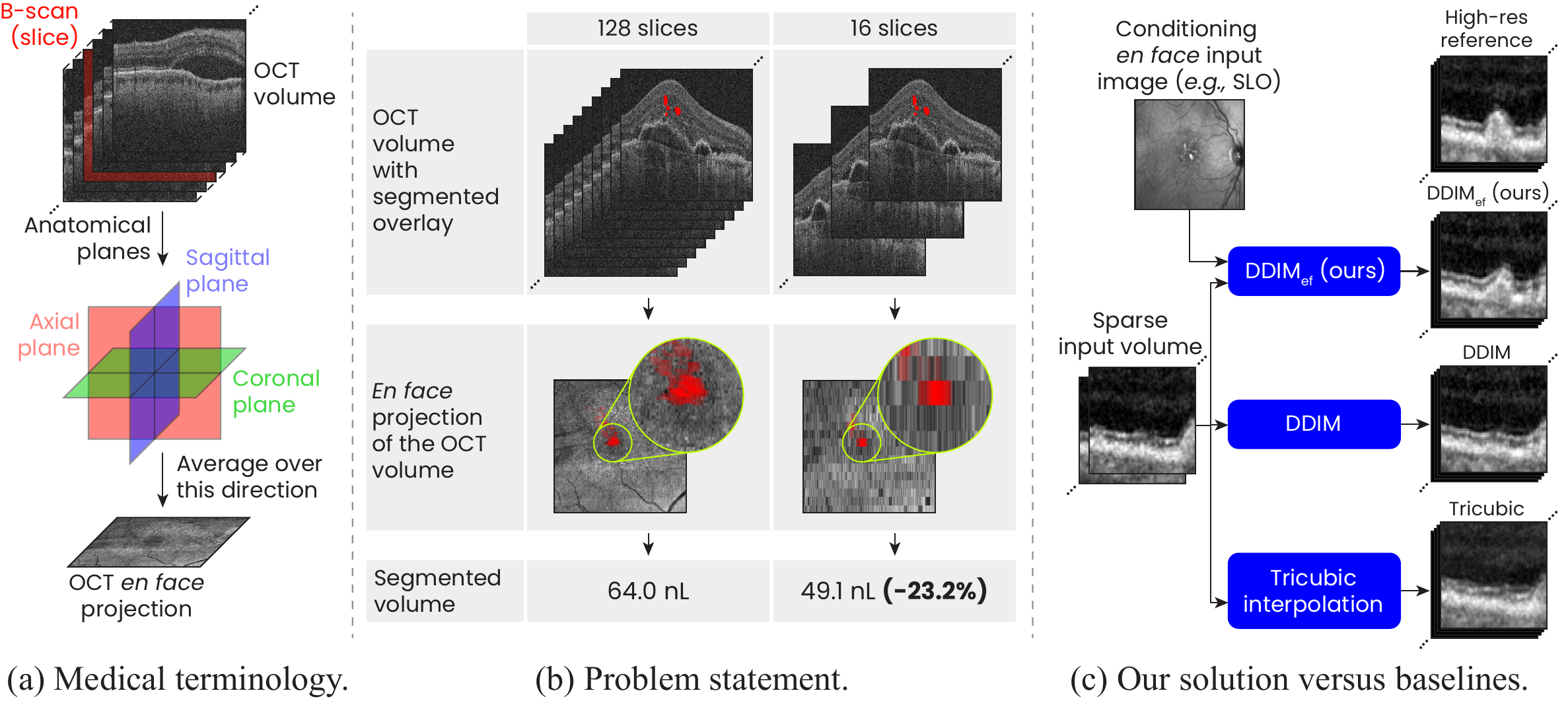}
    \caption{Overview of our proposed approach and background. (a) The medical terminology related to OCT volumes. (b) Low slice densities in volumetric images can lead to inaccurate quantifications. (c) We propose to upsample the number of slices by conditioning a diffusion model with \enface{} images. In our experiments, we use SLO as conditioning \enface{} data but our method could be extended to other types. In the shown example, our model is the only approach that correctly generates the druse (the bright bump in the retina).}
\label{fig:diffusion:overview_background_and_method}
\end{figure*}

A prominent example of a modality often affected by this is optical coherence tomography (OCT). OCT is commonly acquired as a raster, where multiple line scans generate B-scans (slices), and multiple slices generate a volume (see Fig. \ref{fig:diffusion:overview_background_and_method}a). The spacing between slices can vary substantially between OCT devices and imaging protocols \cite{Wu23,Bogu19,Fars14,Garz22}. 
This can hamper consistent biomarker quantification \cite{Vela17,Bara12,Schm22,Once22,Nitt11}. 
Fig. \ref{fig:diffusion:overview_background_and_method}b illustrates this issue by demonstrating fluid volume estimations in a single retinal OCT volume for various slice spacings. We can observe a 23.2\% drop in estimated fluid volume when the slice spacing increases by a factor of 8. 

A possible solution to these imprecise measurements is to standardize volumes with low slice density to high-resolution data through reliable super-resolution methods. Several super-resolution approaches have been proposed for OCT \cite{Huan19b,Xu18,Das20,Qiu21} but they all aim to improve the resolution within individual B-scans. 
Approaches to reduce anisotropy have been proposed for other volumetric medical images, such as computed tomography (CT) \cite{Li20b} and magnetic resonance imaging (MRI) \cite{Sand22}. These methods use deep learning models to upsample the number of slices based solely on low-resolution input data during inference.

A major drawback of these works is their lack of knowledge about anatomical and pathological structures that fall between two adjacent slices. This can lead to hallucinating models that generate incorrect biological features, potentially resulting in misdiagnoses or otherwise harmful clinical outcomes.
We hypothesize that including information about regions between slices as input to a super-resolution model helps make better-informed generative decisions that correctly reflect the biological truth.

Therefore, we propose a method based on 3D diffusion models to increase slice density by utilizing additional imaging modalities as conditioning information (see Fig. \ref{fig:diffusion:overview_background_and_method}c). 
We use diffusion models, as they have been shown to outperform other popular generative models such as generative adversarial networks at generating high-quality images \cite{Dhar21}, super-resolution \cite{Saha22}, and leveraging multimodal data as conditioning information \cite{Romb22}. These capabilities align well with the objectives of our study.

We evaluate our developed method on OCT data while conditioning on scanning laser ophthalmoscopy (SLO) fundus images.
SLO is a 2D \enface{} (\ie parallel to the coronal plane) imaging modality that is commonly acquired alongside OCT scans. OCT devices internally use SLO images as a reference to position the OCT acquisition at the desired anatomical location \cite{Auma19}.
Our method can be extended to include other modalities, such as color fundus photography (CFP) and fundus autofluorescence (FAF), potentially resulting in even better-informed models.

We hope this approach is a valuable step towards more isotropic, high-quality, and standardized volumetric imaging, allowing for more consistent biological measurements and diagnoses in the future.

\section{Methods}
\begin{figure*}[!t]
    \centerline{\includegraphics[width=\linewidth]{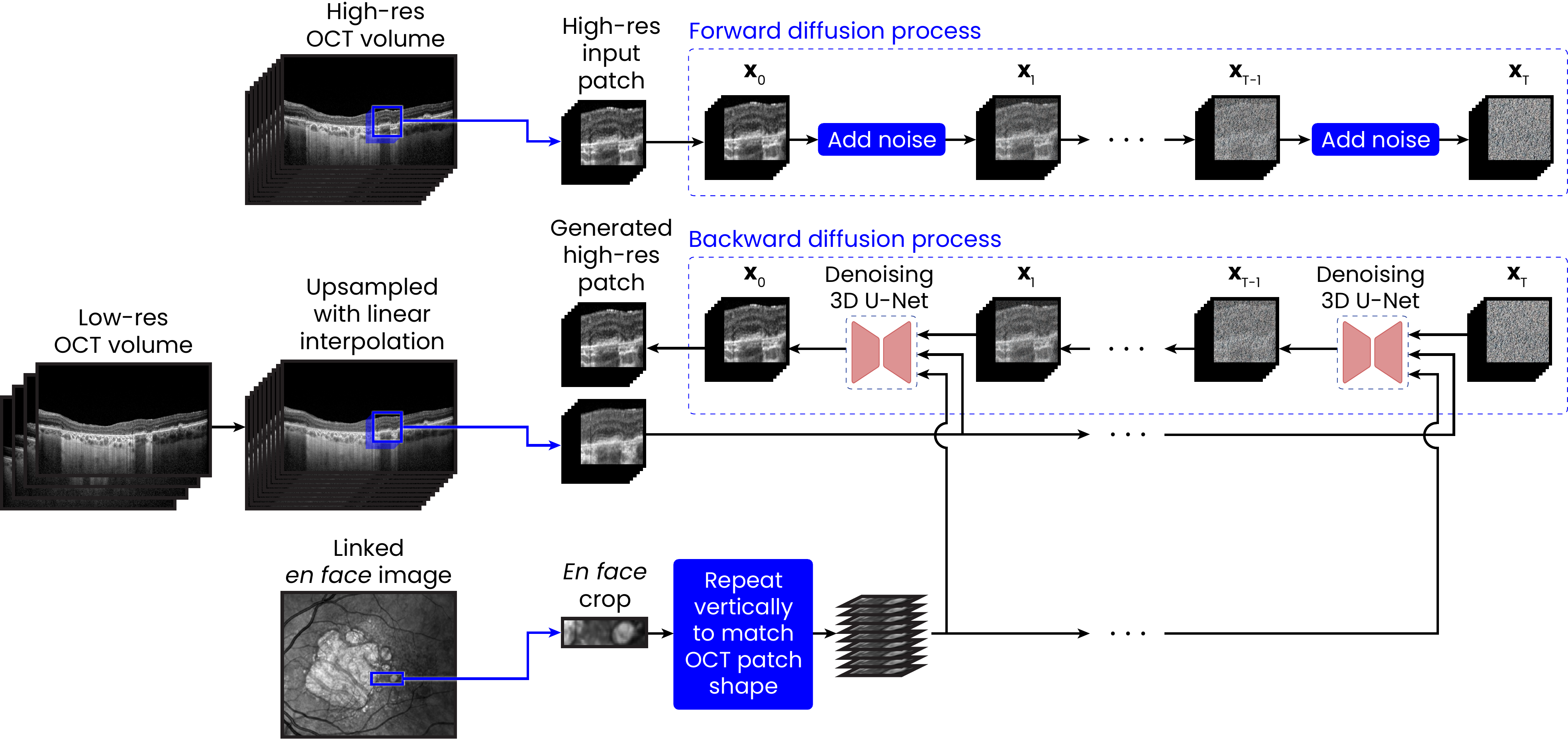}}
    \caption{\Enface{}-conditioned diffusion model overview.}
    \label{fig:diffusion:methods_overview}
\end{figure*}
    
\newcommand{\x}{\mathbf{x}}

Our model is a 3D diffusion model trained to generate high-resolution volumes by upsampling the number of slices (\ie increasing the B-scan density in OCT). It is conditioned on both \enface{} imaging data and the low-resolution counterpart of the high-resolution target (see Fig. \ref{fig:diffusion:methods_overview}). We provide a brief introduction to diffusion models and their associated symbols in Section \ref{sec:diffusion:diffusion_models}. In Section \ref{sec:diffusion:en_face_conditioned_super_resolution}, we describe how we adapt diffusion models for \enface{} conditioned super-resolution. In short, the low-resolution image is concatenated with a reshaped \enface{} image along the channel dimension, which we subsequently input as conditional information to the denoising model. The sampling process, including the use of Denoising Diffusion Implicit Model (DDIM) \cite{Song21} sampling, overlapping patches enabled by RePaint \cite{Lugm22}, and classifier-free guidance (CFG) \cite{Ho22a}, is detailed in Section \ref{sec:diffusion:sampling_process}. Finally, we present the network architecture and implementation details in Section  \ref{sec:diffusion:architecture_and_implementation}.

\subsection{Diffusion models}
\label{sec:diffusion:diffusion_models}

Diffusion models are generative models consisting of a forward diffusion process and a backward diffusion process \cite{Ho20}. In the forward diffusion process, over many timesteps $T$, more and more noise is gradually added to an input image $\x_0$, resulting in noisy images $\x_1, \ldots, \x_T$. This process $q$ can be formulated with a variance schedule $\beta_1,\ldots,\beta_T$ as follows:

\begin{equation}
    q(\x_t|\x_{t-1}) := \mathcal{N}(\x_t;\sqrt{1-\beta_t}\x_{t-1},\beta_t \mathbf{I}).
\end{equation}

As shown by Ho \textit{et al.} \cite{Ho20}, we can directly obtain $\x_t$ given $\x_0$ using the following equation:

\begin{equation}
    \x_t = \sqrt{\bar\alpha_t}\x_0+\sqrt{1-\bar\alpha_t}\boldsymbol\epsilon,
\end{equation}

with $\boldsymbol\epsilon \sim \mathcal{N}(\boldsymbol{0}, \mathbf{I})$, $\alpha_t := 1 - \beta_t$, and $\bar\alpha_t := \prod_{s=1}^t \alpha_s$.

In the backward diffusion process $p_\theta$, $\x_{t-1}$ is predicted for any $t \in \{1, \ldots, T\}$ by denoising $\x_{t}$ using a trained denoising model, optimized by model parameters $\theta$. This model generally uses some variant of the U-Net \cite{Ronn15} architecture (see Section \ref{sec:diffusion:architecture_and_implementation} for the implementation we use). Following Salimans \textit{et al.} \cite{Sali22}, our trained denoising model does not predict the noise $\boldsymbol\epsilon$ or image $\x_0$ directly, but uses $\mathbf{v}$-prediction parameterization as this prevents intensity shifting artifacts in super-resolution models \cite{Ho22}, where $\mathbf{v}_t := \sqrt{\bar\alpha_t} \boldsymbol\epsilon - \sqrt{1-\bar\alpha_t} \x_0$. We use the mean squared error (MSE) loss to train our denoising model $v_\theta$:

\begin{equation}
    \mathcal{L} := || v_\theta(\x_t, t) - \mathbf{v}_t ||_2^2.
\end{equation}

\subsection{\Enface{} conditioned super-resolution}
\label{sec:diffusion:en_face_conditioned_super_resolution}
To enable the use of diffusion models for \enface{} conditioned super-resolution, we combine the approach for generating high-resolution from low-resolution images from SR3 \cite{Saha22} with our proposed method of conditioning the denoising model with \enface{} information. In this paper, we use SLO images but our approach could be extended to other \enface{} data.

Specifically, we adapt the input to the conditional denoising model as follows:

\begin{equation}
    \hat{\mathbf{v}}_t = v_\theta(\x_t, t, \x_{LR}, \x_{enface}),
\end{equation}

where $\hat{\mathbf{v}}_t$ is the output of the denoising model, $\x_t$ is the noisy image at timestep $t$, $\x_{LR}$ is the corresponding low-resolution image upsampled to match the dimensions of $\x_t$ using linear interpolation, and $\x_{enface}$ is the \enface{} image.

$\x_{LR}$ and $\x_{enface}$ are fed similarly into the denoising model. Following SR3 \cite{Saha22}, we concatenate the noisy input $\x_t$ with the low-resolution image $\x_{LR}$ along the channel dimension. Since \enface{} images generally do not align with their corresponding OCT scans in the \enface{} plane, registration of these two images is required (see Section \ref{sec:diffusion:appendix:additional_dataset_details}). This registration step results in spatial correspondence and ensures the \enface{} image has the same shape in the \enface{} plane as the OCT volume. To reach an image with the same 3D shape as $\x_t$ and $\x_{LR}$, we repeat the registered 2D $\x_{enface}$ image in the y-direction $H$ times, where $H$ is the height of $\x_t$. This 3D tensor is concatenated together with $\x_t$ and $\x_{LR}$ along the channel dimension. We subsequently input the resulting tensor into the model.

Due to computational limitations, we work with image crops for $\x_t$, $\x_{LR}$, and $\x_{enface}$. These crops all correspond in terms of their location and size. No noise is added to $\x_{LR}$ and $\x_{enface}$ for any timestep $t$.

\subsection{Sampling process}
\label{sec:diffusion:sampling_process}
During sampling, there are a few key distinctions in the processing pipeline compared to the backward diffusion process during training. Firstly, we use DDIM \cite{Song21} sampling, which allows for accelerated sampling by reducing the number of timesteps while sampling.

Secondly, we train our denoising model with patches, but we are interested in generating full high-resolution volumes during sampling. To prevent artifacts near the borders of patches, we use overlapping patches. We use RePaint \cite{Lugm22} to facilitate this overlapping strategy, an image inpainting approach for diffusion models (see Section \ref{sec:diffusion:appendix:overlapping_patches} for more details). 

Thirdly, to minimize image artifacts showing an overall intensity in the generated slices that is different from the overall intensity in the slices already existing in the low-resolution volume, we implemented a post-processing normalization step. In this step, we normalized the mean and standard deviation of the intensities in the generated slices to match those in the slices that already existed in the low-resolution volume. This normalization step was performed separately for each OCT volume.

Lastly, to influence how much the denoising model uses the \enface{} information for its generative decisions, we employ CFG \cite{Ho22a}. In CFG, during training, the conditional information is dropped for a random number of samples in each batch with some probability $p_{uncond}$. In practice, when the conditional information is dropped, we feed an image with all pixels set to zero. This results in a jointly trained conditional denoising model $v_\theta(\x_t, t, \x_{LR}, \x_{enface})$ and unconditional denoising model $v_\theta(\x_t, t, \x_{LR})$. During sampling, we can then linearly combine the conditional and unconditional model predictions using a guidance scale hyperparameter $w$:

\begin{equation}
    \Tilde{v}_\theta(\x_t, t, \x_{LR}, \x_{enface}) = (1 - w)v_\theta(\x_t, t, \x_{LR}) + w v_\theta(\x_t, t, \x_{LR}, \x_{enface}).
\end{equation}

\subsection{Network architecture and implementation details}
\label{sec:diffusion:architecture_and_implementation}
For our denoising network, we use the U-Net architecture described by Pinaya \textit{et al.} \cite{Pina23}, in which the timestep embedding is integrated into the model via residual connections. The U-Net uses 3D convolutions and has a depth of four U-Net levels with 32, 64, 128, and 256 channels, respectively, in each level with two residual blocks per level. We use self-attention at the deepest U-Net level with a single attention head. 
We train using the Adam optimizer with a learning rate of $5\times10^{-5}$, a batch size of 16, for 20~000 epochs. All images were normalized between -1 and 1 before cropping.

In our experiments, the number of timesteps for the diffusion process was $T=1000$ during training with a \textsc{scaled linear} \cite{Pina23} schedule and a $\beta_t$ range of 0.0005 to 0.0195. We employed DDIM sampling with 100 timesteps, resulting in a 10$\times$ time efficiency improvement. We do not perform any resampling steps when inpainting using RePaint \cite{Lugm22} to allow for faster sampling. We use the MONAI generative AI framework \cite{Pina23} for implementing our diffusion model.

During training, we use 3D crops of size 128 $\times$ 128 $\times$ 16 for $\x_t$ and $\x_{LR}$, and a 2D crop of size $128 \times 1 \times 16$ for $\x_{enface}$. During sampling, we use a crop size of 496 $\times$ 496 $\times$ 16 for $\x_t$ and $\x_{LR}$, and a 2D crop of size 496 $\times$ 1 $\times$ 16 for $\x_{enface}$. When sampling full high-resolution volumes, we used an overlap of 25\%, 25\%, and 50\% for the x-, y-, and z-direction, respectively. During training, we prepared our crops such that each 5\textsuperscript{th} and each 13\textsuperscript{th} slice in $\x_{LR}$ were identical to the 5\textsuperscript{th} and 13\textsuperscript{th} slice in $\x_0$, respectively. The other slices were interpreted using linear interpolation.

\subsection{Data}
\label{sec:diffusion:macustar_dataset}

For training and evaluating our diffusion models, we used OCT volumes and corresponding SLO images from the MACUSTAR study \cite{Fing19}. MACUSTAR is a clinical study on age-related macular degeneration (AMD) that is carried out across 20 sites in 7 European countries. The dataset from this study contains data from patients with varying disease severities (no, early, intermediate, and advanced AMD).

We used the patient data from the cross-sectional part of the MACUSTAR study and only included Heidelberg Spectralis OCTs with at least 237 B-scans. This resulted in a total set of 302 patients. We randomly split this set of patients in 181 (60\%), 60 (20\%), and 61 (20\%) patients for training, validation, and testing, respectively. As multiple OCT volumes were available for each patient (from both eyes and either one, two, or three visits in the cross-sectional study) this resulted in 721 and 236 OCT volumes for the training and validation set, respectively. For the test set, we only used the OCT volume from the study eye, defined for the MACUSTAR study, from the first visit, resulting in 61 OCT volumes. More details regarding the dataset and pre-processing can be found in Appendix \ref{sec:diffusion:appendix:additional_dataset_details}.

\section{Experiments}

\newcommand{\DDIM}{DDIM\xspace}
\newcommand{\DDIMefnocfg}{DDIM\textsubscript{ef} (no CFG)\xspace}
\newcommand{\DDIMef}{DDIM\textsubscript{ef}\xspace}

\newcommand{\LPIPSaxi}{LPIPS\textsubscript{axi}\xspace}
\newcommand{\LPIPScor}{LPIPS\textsubscript{cor}\xspace}
\newcommand{\LPIPSsag}{LPIPS\textsubscript{sag}\xspace}
\newcommand{\LPIPSefproj}{LPIPS\textsubscript{efproj}\xspace}
\newcommand{\LPIPStwohalfD}{LPIPS\textsubscript{2.5D}\xspace}

\label{sec:diffusion:experiments}

We evaluate our approach for the task of upsampling the number of slices in the image volume with a factor of 8. For the sake of simplicity, when generating full volumes as described in Section \ref{sec:diffusion:sampling_process}, we drop the last slice in the OCT volume in the test set, resulting in OCT volumes with 240 instead of 241 slices. Hence, we upsample low-resolution volumes with 30 slices to high-resolution volumes with 240 slices. The resolution of individual slices was not changed.

We refer to our proposed diffusion model with \enface{} conditioning and CFG with guidance scale $w = 2$ as \DDIMef. To measure the effect of \enface{} conditioning and CFG, we compare \DDIMef~with the two models: \DDIMefnocfg, and \DDIM. \DDIM is the proposed approach with \enface{} conditioning turned off during sampling. \Enface{} conditioning is turned off by feeding an image with all pixels set to zero as the conditional image, which is also done during training with a probability of $p_{uncond}$ to enable CFG (see Section \ref{sec:diffusion:sampling_process}). The proposed model with \enface{} conditioning, but without CFG, will be referred to as \DDIMefnocfg.  Furthermore, we compare these methods with the more traditional upsampling method of tricubic interpolation.

\subsection{Evaluation}
\label{sec:diffusion:experiments_evaluation}

We report the classical image similarity metrics MSE, SSIM \cite{Wang04}, and PSNR, computed between the $8 \times$ upsampled low-resolution images using DDIMs and tricubic interpolation, and the high-resolution reference images.
As noted by Saharia \textit{et al.} \cite{Saha22}, these classical metrics may not be optimal for evaluating super-resolution methods, as they were shown to correlate poorly with human perception and heavily penalize synthetic high-frequency details that deviate from the reference, favoring than blurry images instead.

Therefore, we also evaluate with Learned Perceptual Image Patch Similarity (LPIPS) \cite{Zhan18a}.
We used the LPIPS metric implementation provided by the authors of the original LPIPS paper \cite{Zhan18a} for their metric based on an ImageNet pre-trained AlexNet \cite{Kriz12}. Since this evaluation method was developed for 2D images, we modified it for 3D data by calculating the LPIPS metric on all 2D slices in the axial, coronal and sagittal planes in the volume, resulting in the metrics \LPIPSaxi, \LPIPScor, and \LPIPSsag, respectively. We also report \LPIPStwohalfD, which is the average of these three metrics. This approach for 3D data is available in the publicly available implementation from the MONAI generative AI framework \cite{Pina23}. Additionally, we calculated \LPIPSefproj, which compares two OCT \enface{} projections generated by averaging all columns in each slice of the volume, resulting in a 2D \enface{} image.

Even though LPIPS may be considered a more suitable evaluation approach than the classical evaluation metrics, it is not guaranteed to consider the structures of interest enough, since the underlying model was only trained on natural images. Therefore, we also present qualitative examples.

\section{Results}
\begin{table}[!t]
    \centering
    \caption{Classical image similarity metrics (MSE, SSIM, and PSNR) and perceptual metrics (all LPIPS variants) calculated on the test set between the original high-resolution OCT volumes and low-resolution images that were 8$\times$ upsampled in the slice-direction using tricubic interpolation and our proposed DDIM methods. Results are presented as the mean $\pm$ standard deviation over all OCT volumes. Bolded values indicate the best values in terms of the mean performance.}
    \label{tab:diffusion:quantitative}
    \begin{tabular}{@{}lrrrrr@{}}

    \toprule
     & Tricubic & DDIM & DDIM\textsubscript{ef} (no CFG) & DDIM\textsubscript{ef} \\ \midrule
    MSE $\downarrow$ & \textbf{0.006\footnotesize{ $\pm$ 0.002}}& \textbf{0.006\footnotesize{ $\pm$ 0.003}}& \textbf{0.006\footnotesize{ $\pm$ 0.003}}& \textbf{0.006\footnotesize{ $\pm$ 0.003}} \\ 
    SSIM $\uparrow$ & \textbf{0.451\footnotesize{ $\pm$ 0.116}}& 0.444\footnotesize{ $\pm$ 0.107}& 0.447\footnotesize{ $\pm$ 0.107}& 0.447\footnotesize{ $\pm$ 0.107} \\ 
    PSNR (dB) $\uparrow$ & 22.472\footnotesize{ $\pm$ 1.418}& 22.401\footnotesize{ $\pm$ 1.644}& \textbf{22.495\footnotesize{ $\pm$ 1.673}}& 22.450\footnotesize{ $\pm$ 1.683} \\ \midrule
    LPIPS\textsubscript{axi} $\downarrow$ & \textbf{0.120\footnotesize{ $\pm$ 0.027}}& 0.138\footnotesize{ $\pm$ 0.030}& 0.138\footnotesize{ $\pm$ 0.030}& 0.141\footnotesize{ $\pm$ 0.031} \\ 
    LPIPS\textsubscript{cor} $\downarrow$ & 0.548\footnotesize{ $\pm$ 0.103}& \textbf{0.158\footnotesize{ $\pm$ 0.047}}& \textbf{0.158\footnotesize{ $\pm$ 0.048}}& 0.162\footnotesize{ $\pm$ 0.050} \\ 
    LPIPS\textsubscript{sag} $\downarrow$ & 0.540\footnotesize{ $\pm$ 0.088}& \textbf{0.144\footnotesize{ $\pm$ 0.049}}& \textbf{0.144\footnotesize{ $\pm$ 0.049}}& 0.147\footnotesize{ $\pm$ 0.050} \\ 
    LPIPS\textsubscript{2.5D} $\downarrow$ & 0.403\footnotesize{ $\pm$ 0.072}& \textbf{0.147\footnotesize{ $\pm$ 0.041}}& \textbf{0.147\footnotesize{ $\pm$ 0.042}}& 0.150\footnotesize{ $\pm$ 0.043} \\ 
    LPIPS\textsubscript{efproj} $\downarrow$ & 0.231\footnotesize{ $\pm$ 0.055}& 0.063\footnotesize{ $\pm$ 0.039}& \textbf{0.060\footnotesize{ $\pm$ 0.039}}& 0.064\footnotesize{ $\pm$ 0.039} \\ 
    \bottomrule

    \end{tabular}
    \end{table}
    
    \begin{figure*}[!t]
    \centerline{\includegraphics[width=.9\linewidth]{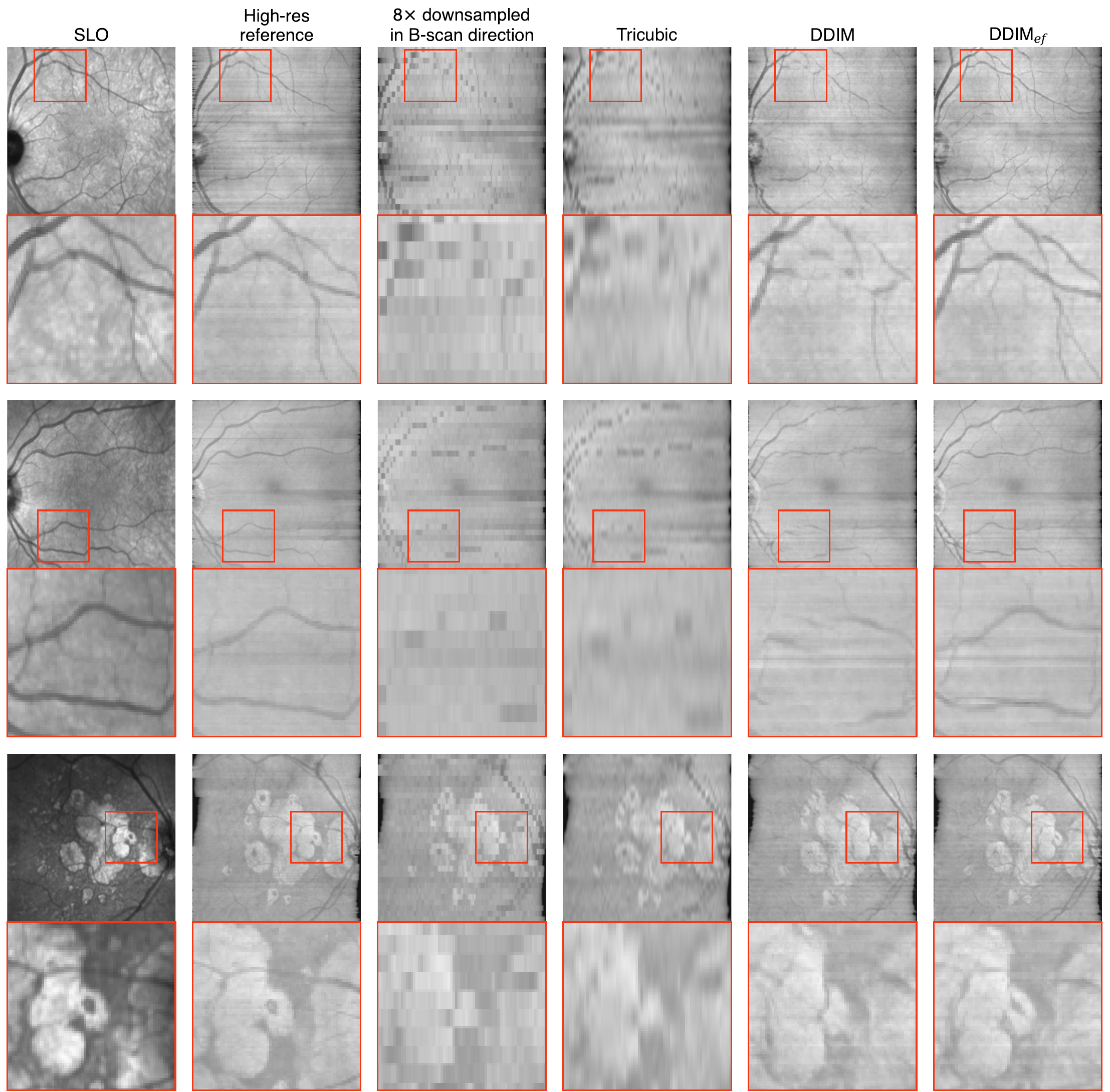}}
    \caption{Examples of \enface{} projections of the OCT volumes generated using tricubic interpolation, the unconditional diffusion model \DDIM, and our proposed \enface{} conditioned diffusion model \DDIMef. These three projections are shown in the last three columns and were all generated from the 8 $\times$ downsampled (in the B-scan direction) volume as input, which is shown in the third column. The first and second rows show the corresponding high-resolution (high-res) reference and scanning laser ophthalmoscopy (SLO) image, respectively. A separate example from a different test set patient is shown in each row. The top images in each row show the full image. Zoomed-in versions of the image crops (red boxes) are shown at the bottom of each row.}
    \label{fig:diffusion:enface_a}
    \end{figure*}

    \begin{figure*}[!ht]
    \centering
    \begin{subfigure}{.8\textwidth}
        \includegraphics[width=\textwidth]{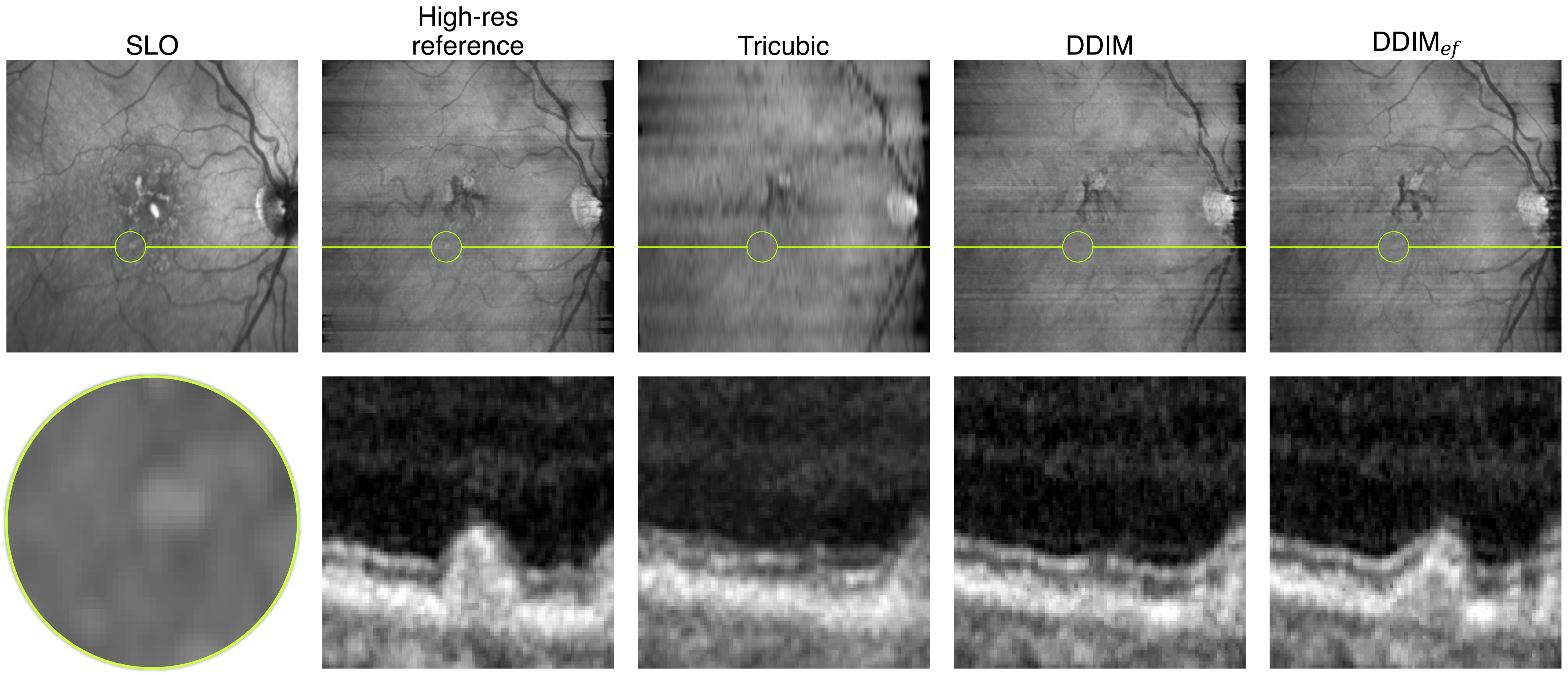}
        \caption{}
        \label{fig:diffusion:nocond_vs_cfg_a}
    \end{subfigure}
    
    \begin{subfigure}{.8\textwidth}
        \includegraphics[width=\textwidth]{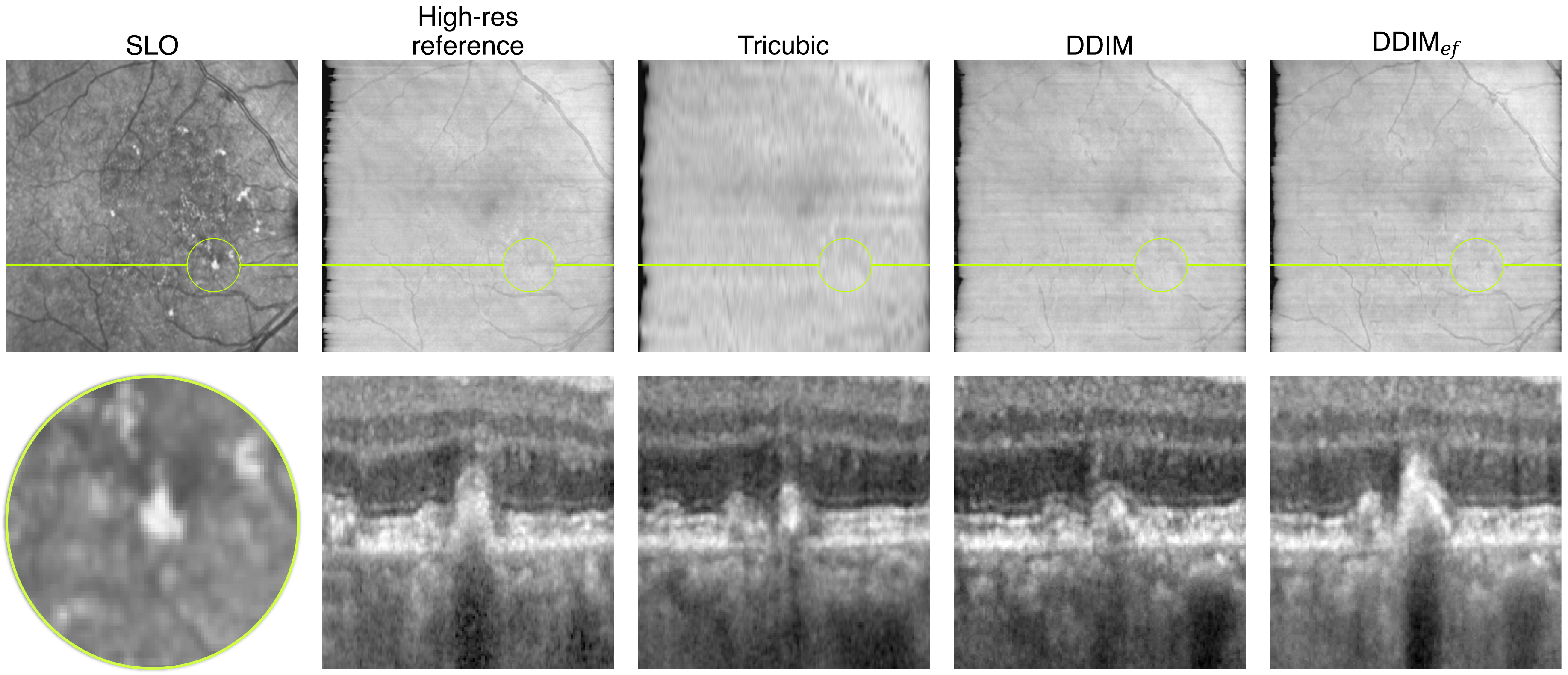}
        \caption{}
        \label{fig:diffusion:nocond_vs_cfg_b}
    \end{subfigure}
    \caption{Examples showing the effect of using \DDIMef, compared to leaving out the \enface{} information (\DDIM), and tricubic interpolation. In the top rows, the SLO image is shown on the left, followed by the \enface{} projections of each OCT volume. The bottom rows show zoomed-in crops. The crop locations are indicated in the top row with a lime circle. The lime horizontal line corresponds to the B-scan location from the shown crops. (a) \DDIMef reconstructs a druse that is present in the high-resolution reference, while it is missing in the OCT crops from tricubic interpolation and \DDIM. The druse also seems to be subtly visible in the SLO image. (b) The large lesion in the center of the OCT crops are more similar to the reference for \DDIMef than for tricubic interpolation and \DDIM. It is also visible in the SLO images as a hyper-intense lesion.}
    \label{fig:diffusion:nocond_vs_cfg}
\end{figure*}

\begin{figure*}[!ht]
\centering

\includegraphics[width=.8\textwidth]{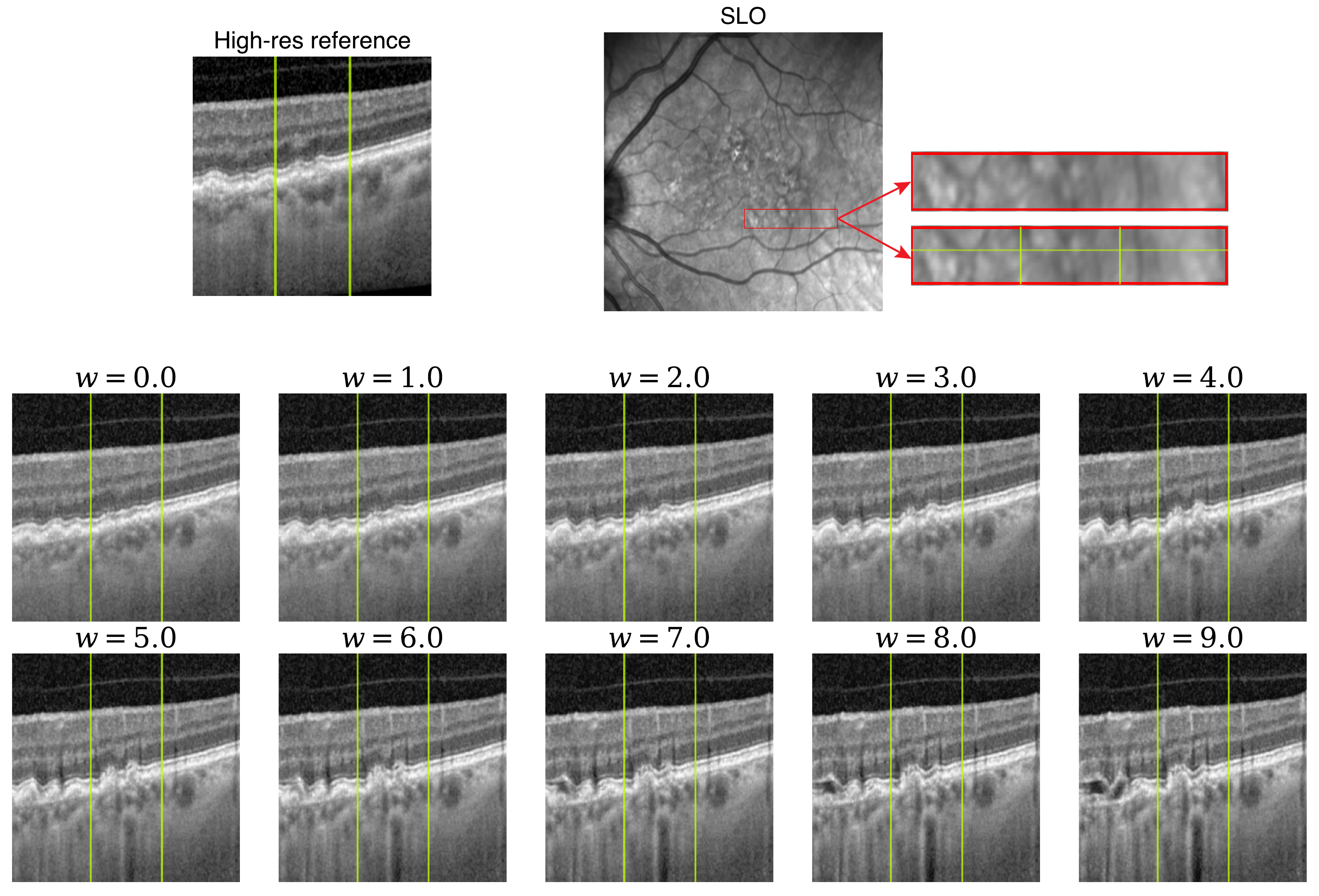}
\caption{The effect of CFG and increasing the guidance scale $w$. In the top left, the high-resolution reference is shown. In the top right, the \enface{} SLO image is shown, in which the crop location is indicated with a red box. The vertical lime lines correspond to the same physical locations throughout the figure. The horizontal lime line in the top right indicates the B-scan location of the shown OCT crops. The bottom two rows show the results from our \enface{} conditioned diffusion model with CFG for various guidance scales. $w = 0$ is equivalent to the unconditional \DDIM, $w = 1$ to \DDIMefnocfg, $w = 2$ to \DDIMef, and $w > 2$ is similar to \DDIMef, but with a larger guidance scale. The two drusen between the lime vertical lines seem to grow with a larger guidance scale.
}
\label{fig:diffusion:cfg_guidance_scale_example}
\end{figure*}

The first three rows of Table \ref{tab:diffusion:quantitative} shows the classical image similarity metrics MSE, SSIM \cite{Wang04}, and PSNR, computed between the $8 \times$ upsampled low-resolution images using DDIMs and tricubic interpolation, and the high-resolution reference images. The perceptual metrics based on LPIPS are shown in the last five rows of Table \ref{tab:diffusion:quantitative}.

For the proposed method, ablated methods and the tricubic interpolation method, we present several figures to illustrate the difference in structural similarity to the high-resolution reference images, sharpness, and coherence within the generated volumes. In Fig. \ref{fig:diffusion:enface_a} and \ref{fig:diffusion:enface_b}, we aim to point out these aspects using the \enface{} projections, allowing one to observe the overall structure of the full generated volumes. Fig. \ref{fig:diffusion:nocond_vs_cfg} and Fig. \ref{fig:diffusion:nocond_vs_cfg_continued} show additional examples of these \enface{} projections while highlighting relevant regions, alongside the corresponding image crops from the underlying OCT volume. Further examples are presented in Fig. \ref{fig:diffusion:renders}, which shows 3D renders of generated and reference volumes, and in Fig. \ref{fig:diffusion:bscan_sequences}, which shows crops of consecutive B-scans with difference maps between generated and reference images. 
Fig. \ref{fig:diffusion:difference_images} shows generation examples and difference maps for several randomly picked image crops.
The effect of increasing the guidance scale $w$ from CFG is illustrated in Fig. \ref{fig:diffusion:cfg_guidance_scale_example}. Specifically, a case of an example OCT crop is shown with several drusen that are also visible in the SLO image.

Model training took approximately 9 days on an NVIDIA A100 GPU. Sampling a full volume with 240 B-scans of 768 $\times$ 496 pixels, the most common size in the test set, resulted in 58 patches. On the aforementioned GPU type, for \DDIMef, sampling a whole volume took approximately 46 minutes. For \DDIM and \DDIMefnocfg, this sampling time was about half (approximately 23 minutes), as CFG doubles the number of required forward passes of the denoising model.

\section{Discussion}

We addressed the large variability in anisotropy across OCT scans, which can lead to inconsistent quantifications. We propose a super-resolution approach that uses SLO images to condition 3D super-resolution diffusion models, aiming for better informed image generations that are closer the biological truth. SLO is commonly already acquired alongside OCT scans, ensuring our method often will not require any additional data beyond what is already available in clinical practice. Furthermore, SLO acquisition is relatively fast, while OCT acquisition time increases with every additional line to be acquired, potentially speeding up the overall acquisition process.

Our qualitative results indicate that our approach can upsample the number of B-scans in OCT volumes by a factor of 8 while improving similarity to high-resolution reference images and overall coherence, compared to a diffusion-based approach without \enface{} information as conditional input. Our method specifically improves the reconstruction of superficial blood vessels and geographic atrophy (see Fig. \ref{fig:diffusion:enface_a} and \ref{fig:diffusion:enface_b}). 
Furthermore, our diffusion models demonstrate visually sharper images than tricubic interpolation.

In terms of the classical image similarity metrics MSE, SSIM, and PSNR, which are suboptimal for evaluating super-resolution methods \cite{Saha22}, tricubic interpolation performed roughly the same as our diffusion models.
In terms of \LPIPScor, \LPIPSsag, and \LPIPSefproj, our diffusion models outperformed tricubic interpolation but slightly underperformed in terms of \LPIPSaxi. 
This finding is in line with our visual observations in Appendix Section \ref{sec:diffusion:appendix:lpips_different anatomical planes}.

In terms of all reported quantitative metrics, using CFG to guide the diffusion model more towards the information in the \enface{} image either slightly decreased performance, or showed no effect (see Table \ref{tab:diffusion:quantitative}). Paradoxically, we found using CFG could lead to structural features that more closely resembled those in the high-resolution reference than when CFG was not employed. An example of this effect is shown in Fig. \ref{fig:diffusion:cfg_guidance_scale_example}. Besides, this example shows that setting the guidance scale $w$ too high can lead to exaggerated structural features (\eg too large drusen) and artifacts.

This study has limitations. (1) Our DDIM models can sometimes introduce imaging artifacts (see Appendix Section \ref{sec:diffusion:appendix:generated_imaging_artifacts}). 
(2) When visually inspecting the dataset, we found that the registration information between OCT B-scans and the SLO image was not always perfect, possibly not allowing the model to utilize the SLO/OCT mapping well sometimes. An improved registration strategy will likely improve generation results and learning speed.
(3) Although we present various qualitative results, we only evaluated our approach quantitatively using image similarity metrics. 
Therefore, we cannot draw conclusions about whether our super-resolution approach improves biomarker quantification. 
This requires a vast amount of manual annotation labour or a reliable segmentation model for the type of OCT data used in this work. As we did not have access to these resources, we leave this evaluation to future work.
(4) Hallucination is a large risk of most generative models, including diffusion models \cite{Kim24}. 
In the context of medical imaging, generative models like ours risk the generation of non-existent lesions (leading to false positives), inflating them (leading to over-quantification), removing them (leading to false negatives), or shrinking them (leading to under-quantification).
Although our \enface{} conditioning mechanism may reduce hallucinations by providing more context to make well-informed generative decisions, sufficient empirical evidence of our method completely preventing this is lacking. 
(5) The sampling time for our diffusion model is relatively long. 
However, approaches exist to reduce this sampling time \cite{Sali22,Romb22}.

We only explored the effect of conditioning OCT super-resolution diffusion models with near-infrared SLO images. Future work could include more \enface{} modalities as conditional information, such as CFP and FAF. As images from those modalities would likely provide additional information than SLO images, we expect this could lead to more accurate super-resolution models. Other metadata, such as functional vision exam data and OCT scans from other devices or protocols, may contain even more useful information.

In our current implementation, we resize and vertically repeat the SLO image patch, enabling concatenation in the channel dimension with the OCT volume. This turns the 2D SLO image into a volume that is processed by 3D convolutions. This is computationally inefficient and the initial resizing can lead to information loss. Future work could focus on designing an architecture that more effectively leverages this multimodal data, possibly using a separate encoder for \enface{} images and a cross-attention mechanism to combine the futures from the different encoders.

The approach of using relatively high-resolution 2D images from a certain modality to condition diffusion models for super-resolving 3D data from another modality could potentially be applied in other medical domains. For example, using high-resolution 2D X-ray imaging as conditional information for super-resolution CT or MRI scans may be an interesting future direction.

In conclusion, we have shown the feasibility of conditioning super-resolution diffusion models to reduce anisotropy in volumetric images with additional and readily available image data, enabling well-informed generative decisions. Specifically, we showed this in the context of OCT super-resolution conditioning on \enface{} images.
We think this can be an important next step towards standardized high-quality OCT and other volumetric imaging, leading to more consistent measurements -- obtained from either downstream manual quantifications or machine learning models -- within and across datasets, studies, and clinical practices.
Furthermore, our approach could facilitate the trustworthiness of generative models and their regulatory approval by mitigating the risk of hallucinations compared to uninformed super-resolution models.

\begin{ack}
This project has received funding from the Innovative Medicines Initiative 2 Joint Undertaking under grant agreement No 116076. This Joint Undertaking receives support from the European Union’s Horizon 2020 research and innovation program and EFPIA.  This communication reflects the author’s view and neither IMI nor the European Union and EFPIA are responsible for any use that may be made of the information contained therein.
\end{ack}

{
\small
\bibliographystyle{plain}  
\bibliography{newrefs}
}


\newpage

\appendix
\counterwithin{figure}{section}

\section{Appendix}

\subsection{Additional dataset and pre-processing details}
\label{sec:diffusion:appendix:additional_dataset_details}
For the training and validation set combined, the number of B-scans per OCT volume varied, with 1 volume containing 237 B-scans, 927 volumes containing 241 B-scans, and 29 volumes containing 512 B-scans. The B-scan sizes ranged from 512 $\times$ 496 pixels to 1536 $\times$ 496 pixels, and the physical OCT volume size ranged from 2.9 $\times$ 1.9 $\times$ 2.9 mm\textsuperscript{3} to 9.2 $\times$ 1.9 $\times$ 7.7 mm\textsuperscript{3}. For the test set, all OCT volumes contained 241 B-scans. The B-scan sizes ranged from 512 $\times$ 496 pixels to 1536 $\times$ 496 pixels, and all OCT volumes had a physical size of approximately 8.8 $\times$ 1.9 $\times$ 7.3 mm\textsuperscript{3}.

Next to the previously described set of OCT volumes, we used the near-infrared confocal SLO images, which Heidelberg Spectralis devices acquire alongside the OCT, as \enface{} modality for our conditional diffusion models. The SLO image was registered to the OCT volume according to the physical linkage information between these two images that was provided by the camera software. We subsequently cropped and resized these SLO images to the same width and height as, respectively, the width and depth from their corresponding OCT volumes.

For some OCT volumes in the dataset, we observed that incidentally adjacent B-scans were not correctly aligned vertically. Therefore, during sampling, B-scans were registered vertically using a grid search for the vertical translation amount and MSE as a cost function, based on a flattened representation of the B-scans (collapsed into columns by averaging over the x-axis).

\subsection{Overlapping patches and inpainting}
\label{sec:diffusion:appendix:overlapping_patches}
\begin{figure*}[!htbp]
\centerline{\includegraphics[width=.75\linewidth]{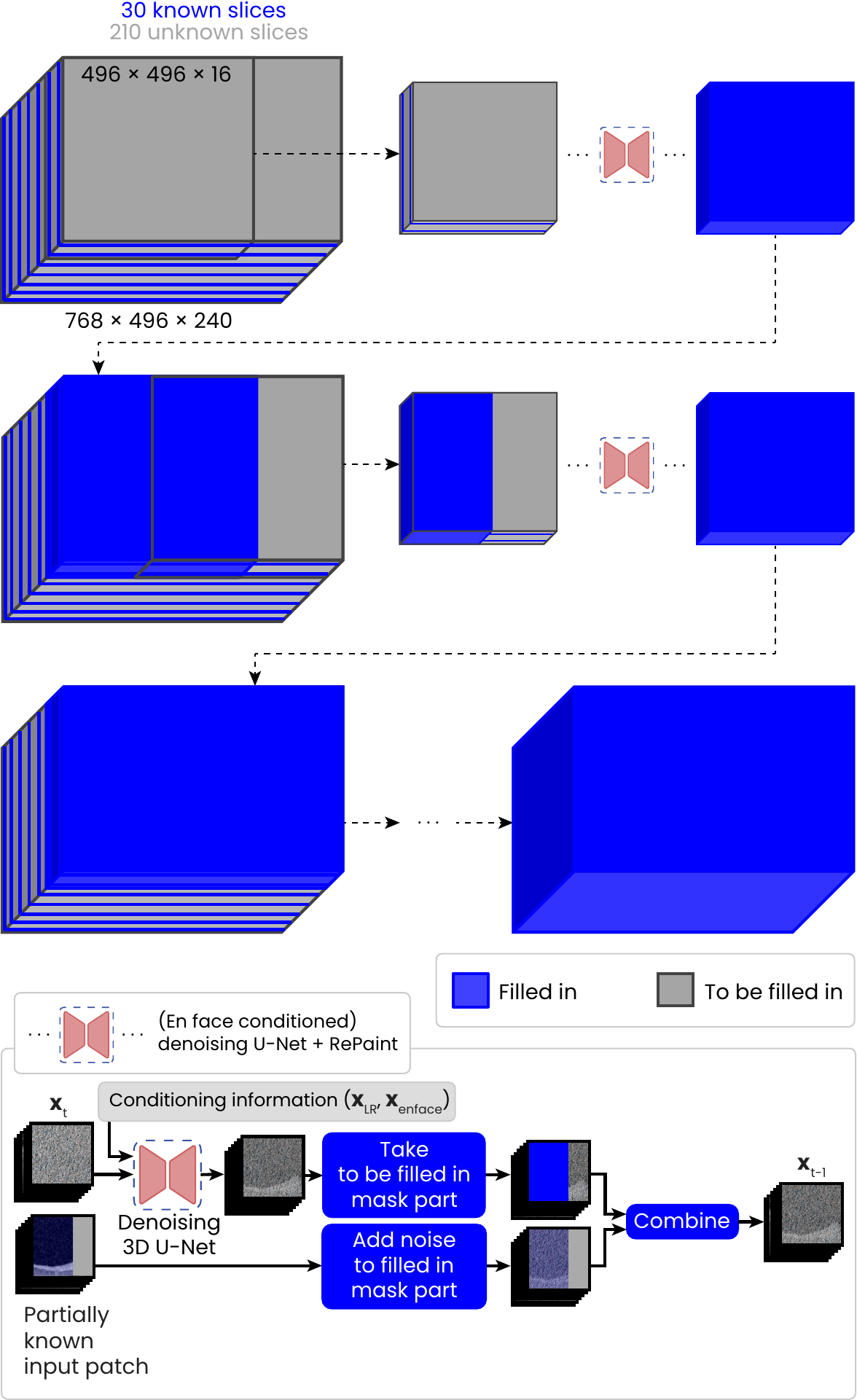}}
\caption{Overview of our overlapping strategy facilitated through RePaint \cite{Lugm22}.}
\label{fig:diffusion:overlapping_patches}
\end{figure*}

During sampling of full volumes, we use overlapping patches, facilitated through inpainting with RePaint \cite{Lugm22}. Inpainting is the task of filling in new content in a specific part of an image, which can be defined by a binary mask. We refer to the image part that needs to be filled in as ``unknown'' and the other part as ``known''. During each timestep $t$ in the sampling process, RePaint combines the ``known'' image part from the input image, which has been noised to the appropriate noise level of timestep $t - 1$, with the ``unknown'' part from the denoised image $\x_{t-1}$ (see the bottom part of Fig. \ref{fig:diffusion:overlapping_patches}). When we generate a complete volume using this patch overlapping strategy, the ``known" region is defined as the area previously generated from an adjacent patch, combined with slices from the original low-resolution image volume (see the top part of Fig. \ref{fig:diffusion:overlapping_patches}). Besides, the patch size during sampling is larger than the one used during training, as we empirically found this to improve the coherence and fidelity of the generated volumes.

\newpage
\subsection{Additional results}

\begin{figure*}[!htbp]
\centerline{\includegraphics[width=\linewidth]{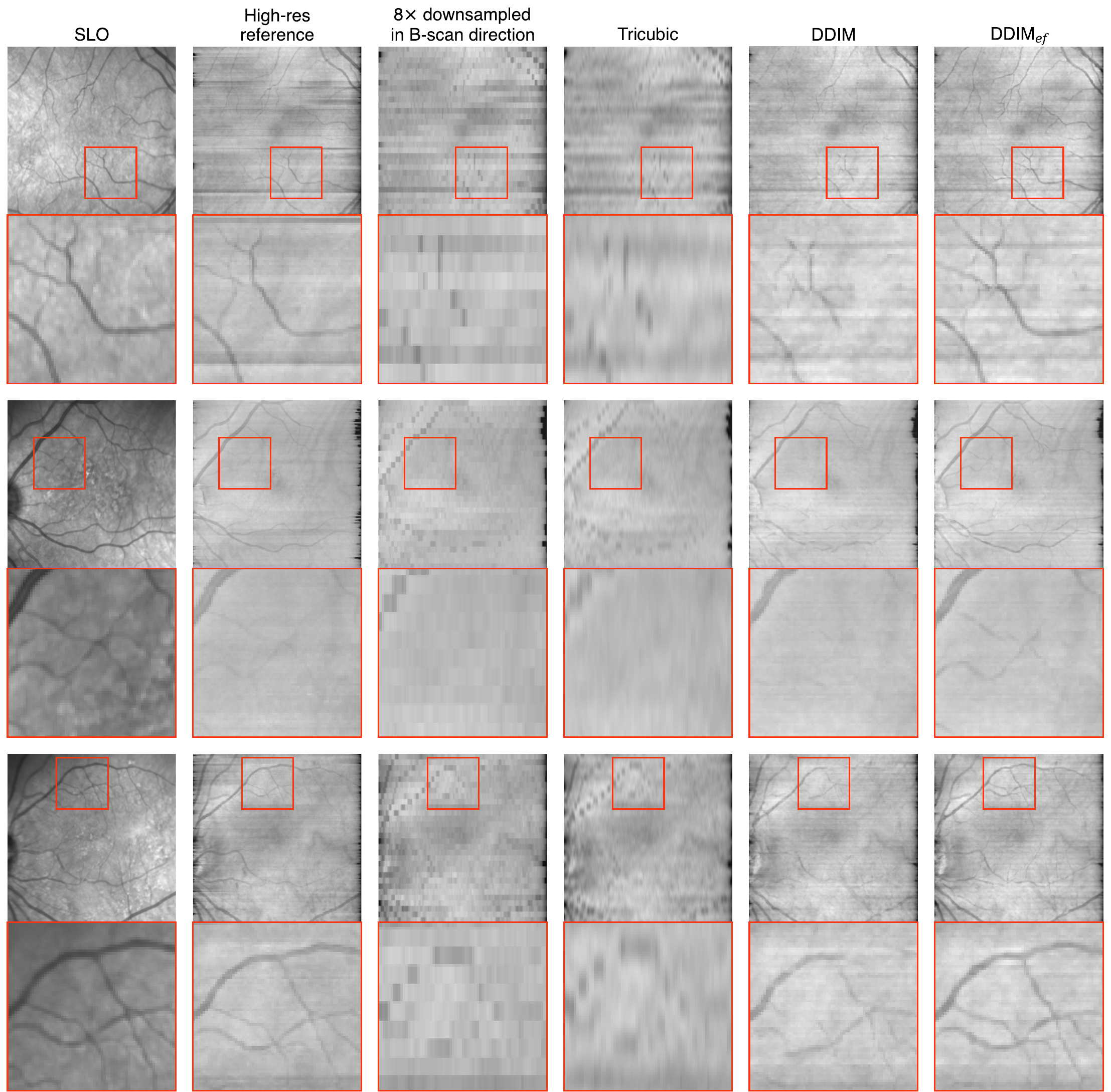}}
\caption{Additional examples of \enface{} projections of the OCT volumes generated using tricubic interpolation, the unconditional diffusion model \DDIM, and our proposed \enface{} conditioned diffusion model \DDIMef (presentation similar to Fig. \ref{fig:diffusion:enface_a}). These three projections are shown in the last three columns and were all generated from the 8 $\times$ downsampled (in the B-scan direction) volume as input, which is shown in the third column. The first and second rows show the corresponding high-resolution (high-res) reference and scanning laser ophthalmoscopy (SLO) image, respectively. A separate example from a different test set patient is shown in each row. The top images in each row show the full image. Zoomed-in versions of the image crops (red boxes) are shown at the bottom of each row.}
\label{fig:diffusion:enface_b}
\end{figure*}

\begin{figure*}[!htbp]
\centering

\begin{subfigure}{\textwidth}
    \includegraphics[width=\textwidth]{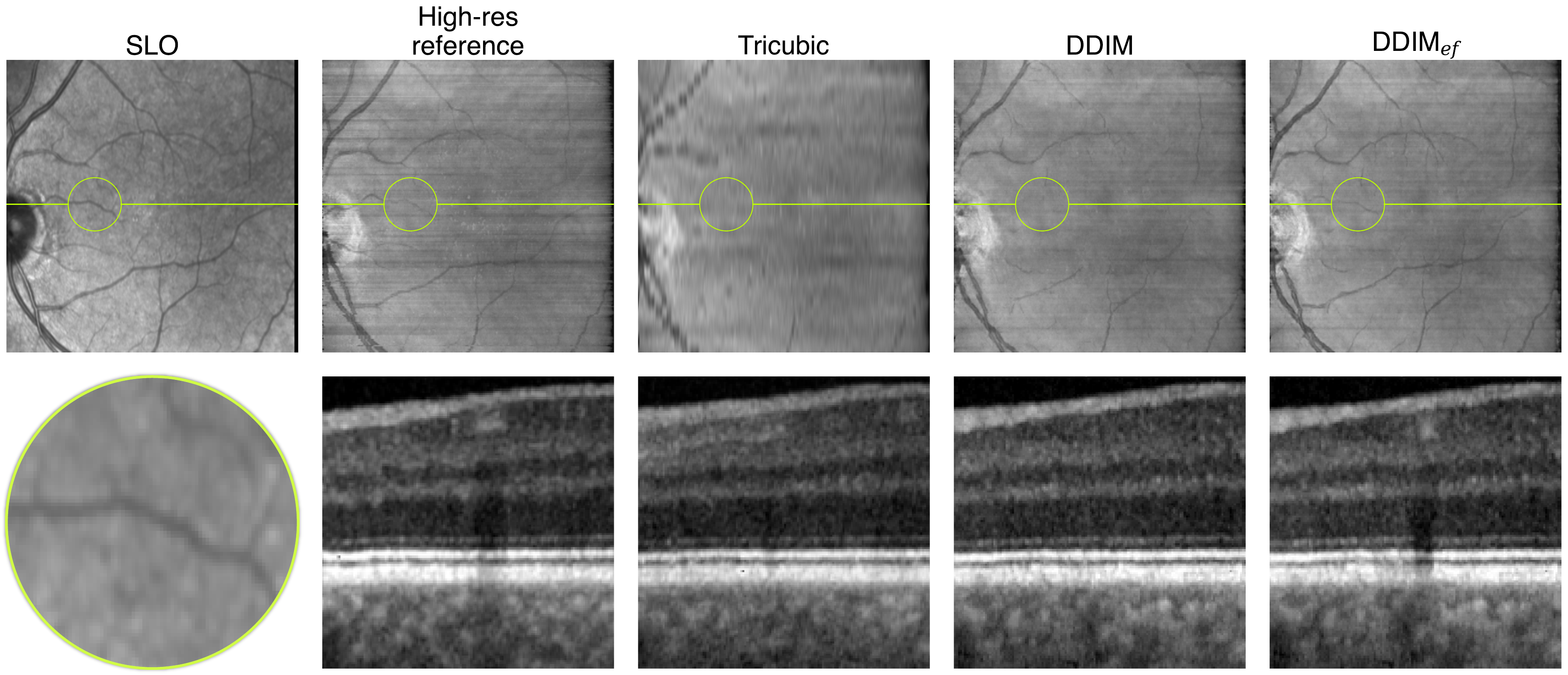}
    \caption{}
    \label{fig:diffusion:nocond_vs_cfg_c}
\end{subfigure}

\begin{subfigure}{\textwidth}
    \includegraphics[width=\textwidth]{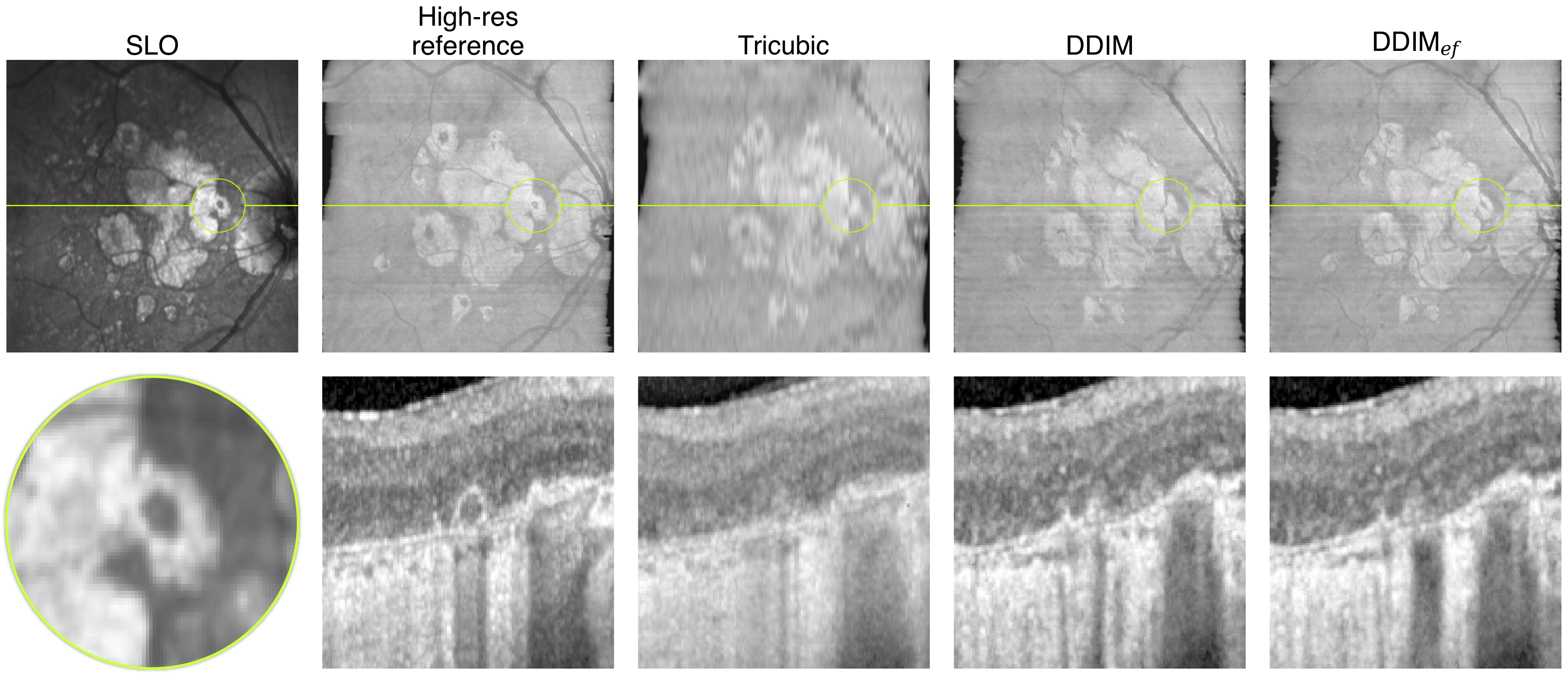}
    \caption{}
    \label{fig:diffusion:nocond_vs_cfg_d}
\end{subfigure}

\caption{Additional examples showing the effect of using \DDIMef, compared to leaving out the \enface{} information (\DDIM), and tricubic interpolation (presentation similar to Fig. \ref{fig:diffusion:nocond_vs_cfg}). In the top rows, the scanning laser ophthalmoscopy (SLO) image is shown on the left, followed by the \enface{} projections of each OCT volume. In the bottom row, zoomed-in crops are shown. The crop locations are indicated in the top row with a lime circle. The lime horizontal line corresponds to the B-scan location from the shown crops. (a) A blood vessel is only reconstructed by \DDIMef. This vessel is also visible in the SLO image. (b) The hypertransmission pattern seems to be best reconstructed by \DDIMef. The \enface{} location of the hypertransmission area is also visible in the SLO image.}

\label{fig:diffusion:nocond_vs_cfg_continued}
\end{figure*}

\begin{figure*}[!htbp]
\centering
\includegraphics[width=\textwidth]{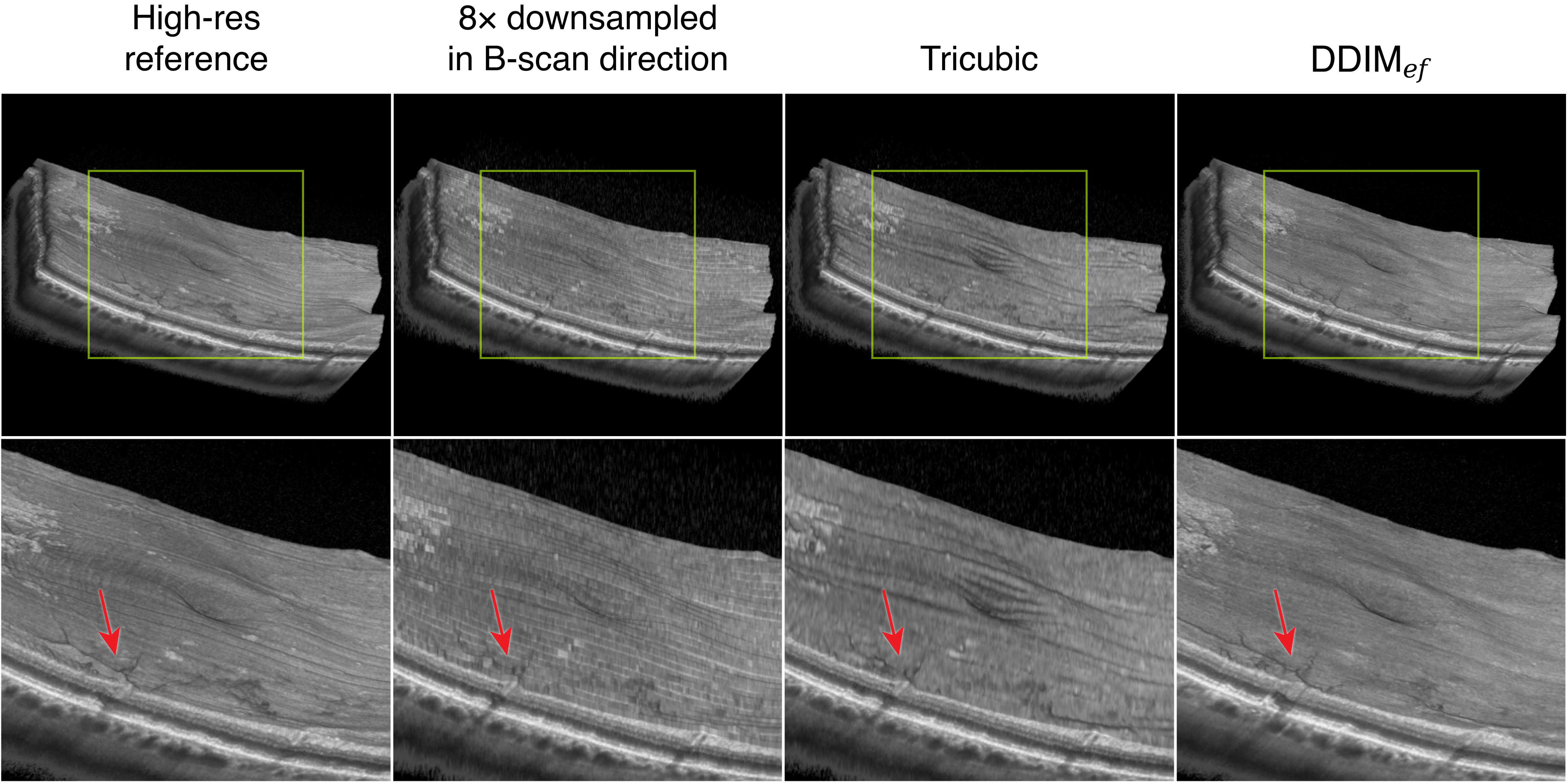}
\caption{3D renders of full OCT volumes, depicted for the high-resolution (high-res) reference, downsampled volume, tricubic interpolated volume, and our proposed method \DDIMef. The bottom row shows zoomed-in versions of the renders in the top row. The lime squares in the top row indicate the zoomed-in area. The red arrows point to a vessel on the inner retina.}  
\label{fig:diffusion:renders}
\end{figure*}

\begin{figure*}[!htbp]
\centering
\includegraphics[width=.82\textwidth]{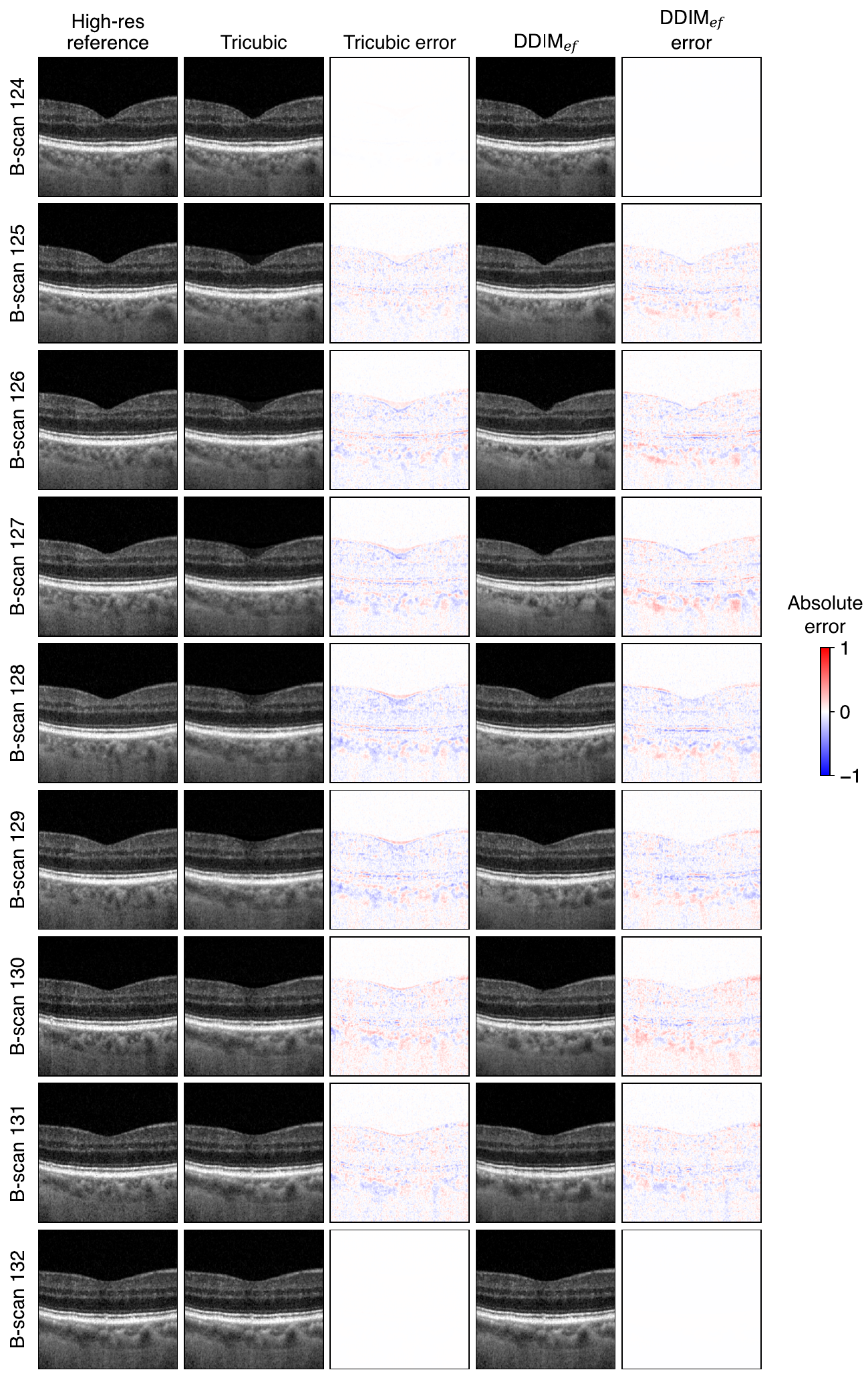}
\caption{An example showing a sequence of B-scans, illustrating that the volume generated with tricubic interpolation is more smoothed than the volume generated by our diffusion model \DDIMef, which is much sharper. This is most evident around slice 128. The first and last shown slices were already present in the low-resolution OCT volume. The third and fifth columns indicate the absolute error.}
\label{fig:diffusion:bscan_sequences}
\end{figure*}

\begin{figure*}[!htbp]
\centering
\includegraphics[width=.82\textwidth]{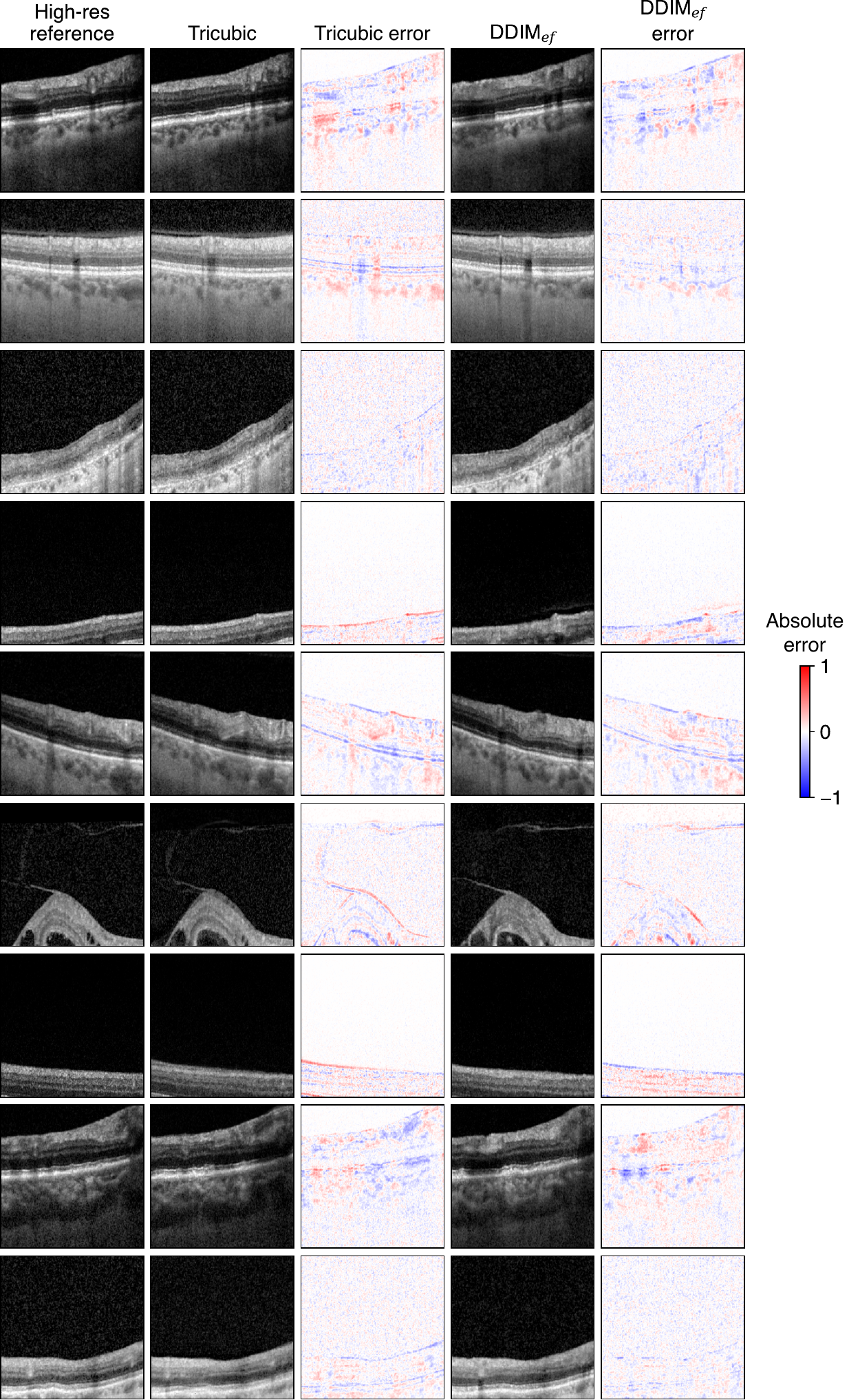}
\caption{Randomly picked 256$\times$256 B-scan crops from ``unknown'' slices in the test set, depicting the origins of absolute errors for tricubic interpolation and \DDIMef. The third and fifth columns indicate the absolute error.}
\label{fig:diffusion:difference_images}
\end{figure*}

\clearpage

\subsection{LPIPS for different anatomical planes}
\label{sec:diffusion:appendix:lpips_different anatomical planes}
\begin{figure*}[!htbp]
\centerline{\includegraphics[width=\linewidth, bb=0 0 800 600]{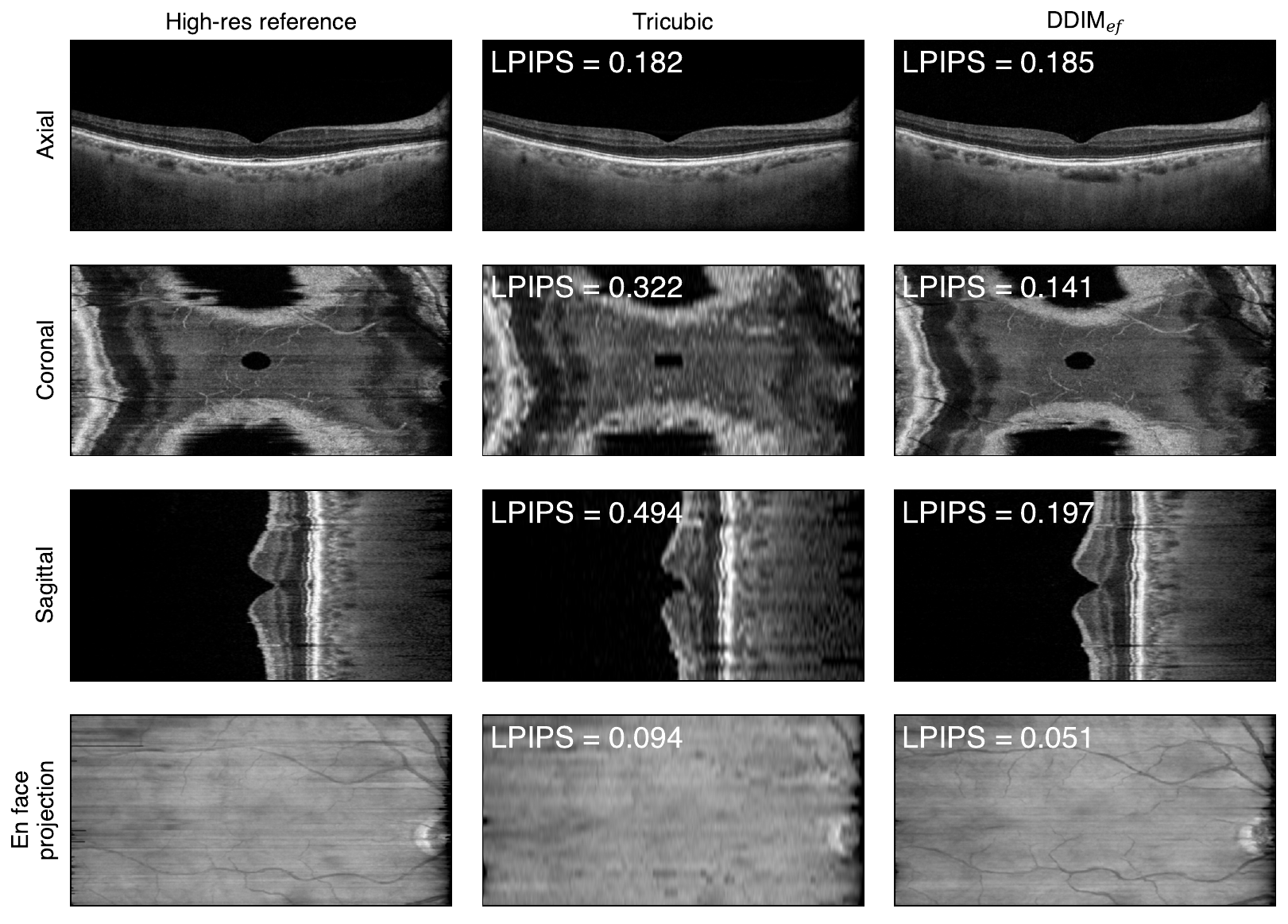}}
\caption{Examples of slices from the three orthogonal anatomical planes and the \enface{} projection image. All shown images originate from the same OCT volume. For tricubic interpolation and \DDIMef, Learned Perceptual Image Patch Similarity (LPIPS) values, computed only using the depicted slices, are shown in the top left corner of each image. For the anatomical planes, the middle slices are shown. For the axial (B-scan) slice, this corresponds to a slice exactly in the middle of two slices that were also present in the low-resolution volume (\ie, the slice was as far away from a ``known'' slice as possible, which is generally a slice that is more difficult to generate accurately than a slice that is closer to a ``known'' slice).}
\label{fig:diffusion:anatomical_planes_example}
\end{figure*}

We found that our diffusion models outperformed interpolation in terms of \LPIPScor, \LPIPSsag, and \LPIPSefproj, but not in terms of \LPIPSaxi.
This finding is in line with the visual observations we make when manually inspecting slices from these anatomical planes. This is illustrated in Fig. \ref{fig:diffusion:anatomical_planes_example}, in which \DDIMef shows higher visual similarity with the high-resolution reference than the image upsampled with tricubic interpolation for the coronal plane, sagittal plane, and the \enface{} projection. For the axial plane, however, we think this is less evident. The LPIPS values, which are also shown in Fig. \ref{fig:diffusion:anatomical_planes_example} for the shown slices, correlate well with these visual observations.

\clearpage
\subsection{Generated imaging artifacts}
\label{sec:diffusion:appendix:generated_imaging_artifacts}
\begin{figure*}[!htbp]
\centering
\begin{subfigure}{.49\textwidth}
\centering
\includegraphics[width=.6\textwidth]{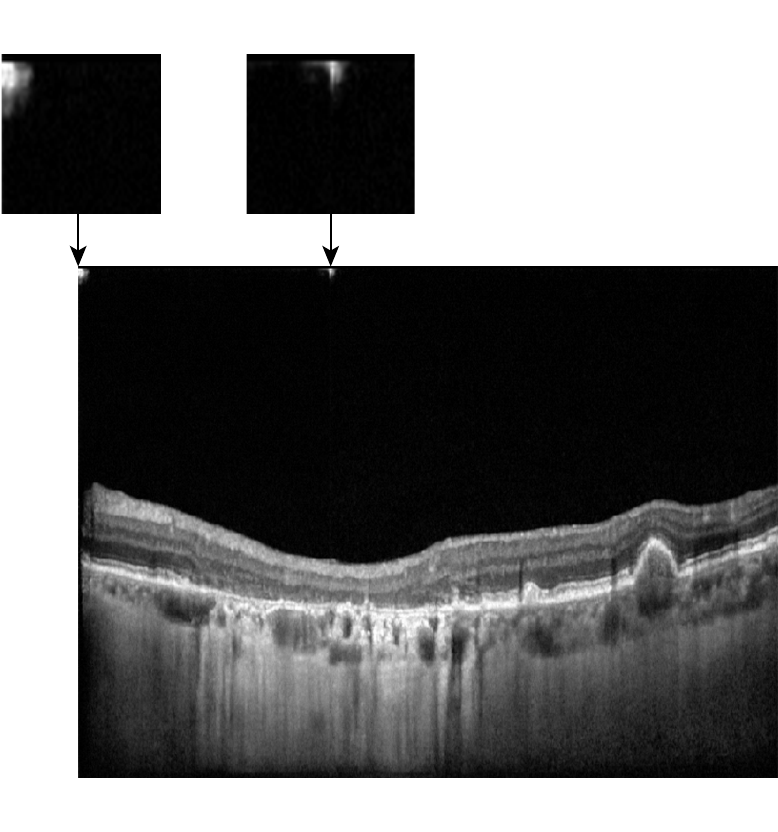}
    \caption{}
    \label{fig:diffusion:failure_mode0}
\end{subfigure}
\begin{subfigure}{.49\textwidth}
\centering
\includegraphics[width=.6\textwidth]{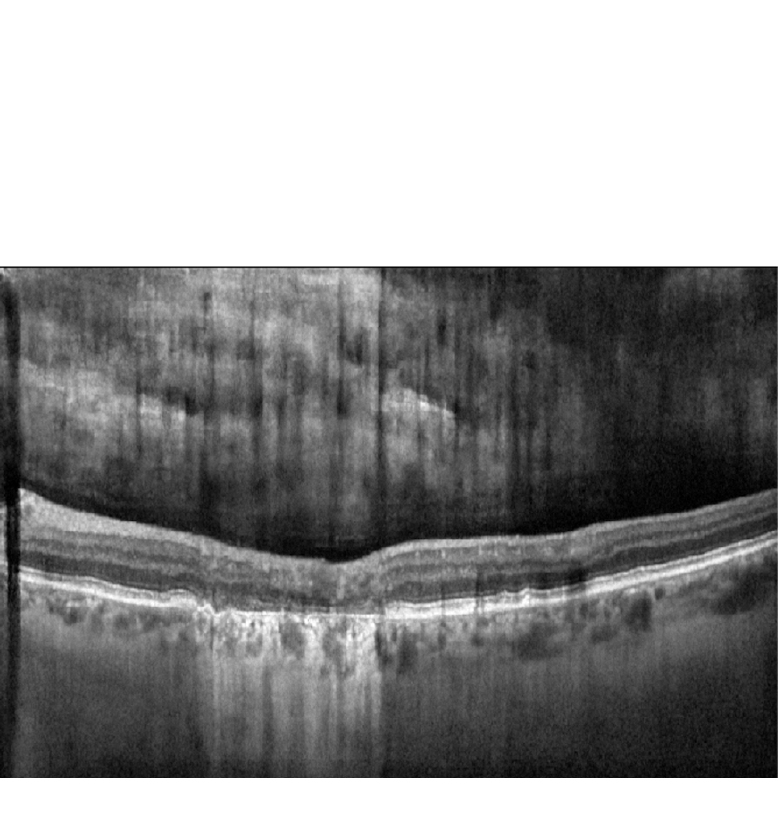}
    \caption{}
    \label{fig:diffusion:failure_mode1}
\end{subfigure}

\caption{Artifact types that sometimes occur in volumes generated by our diffusion models. These artifacts were neither present in the low-resolution input nor the high-resolution reference. (a) Small groups of white pixels in the top left of sampling patches. (b) White areas in the vitreous body. This artifact type mainly seems to occur when two adjacent B-scans in the low-resolution volume were not registered well.}
\label{fig:diffusion:failure_modes}
\end{figure*}

We found that a number of imaging artifacts can occur when generating volumes using our DDIM approach (see \ref{fig:diffusion:failure_modes}).
The occurrence frequency depends on the artifact type. For example, small groups of white pixels are sometimes generated near the top of B-scans (see an example in Fig. \ref{fig:diffusion:failure_mode0}). If they are present, they always seem to be located in the top left of the sampling patches. Furthermore, sometimes very large, bright areas are generated in the vitreous body (see an example in Fig. \ref{fig:diffusion:failure_mode1}). This mainly seems to occur when two adjacent B-scans in the low-resolution volume were not registered well.

\subsection{Ethical approval}
\label{sec:diffusion:appendix:ethical_approval}
Ethical approval of the clinical study data has been obtained, as described in previous works describing the MACUSTAR study \cite{Terh20}.

\ifdefined\macustarConsortium
  \macustarConsortium
\fi

\end{document}